\documentclass[pre,twocolumn]{revtex4-2}
\usepackage{color,graphicx}
\usepackage{epsfig}
\usepackage{amsmath,amssymb}
\usepackage{subcaption}
\usepackage[hidelinks]{hyperref}
\begin{document}
\title{Efficiency of a microscopic heat engine subjected to stochastic resetting}
\author{Sourabh Lahiri}
\affiliation{Department of Physics, Birla Institute of Technology, Mesra, Ranchi, Jharkhand 835215, India}
\author{Shamik Gupta}
\affiliation{Department of Theoretical Physics, Tata Institute of Fundamental Research, Homi Bhabha Road, Mumbai 400005, India}

\begin{abstract}
We explore the thermodynamics of stochastic heat engines in presence of stochastic resetting. The set-up comprises an engine whose working substance is a Brownian particle undergoing overdamped Langevin dynamics in a harmonic potential with a time-dependent stiffness, with the dynamics interrupted at random times with a resetting to a fixed location. The effect of resetting to the potential minimum is shown to enhance the efficiency of the engine, while the output work is shown to have a non-monotonic dependence on the rate of resetting. The resetting events are found to drive the system out of linear response regime even for small differences in the bath temperatures. Shifting the reset point from the potential minimum is observed to reduce the engine efficiency. The experimental set-up for the realization of such an engine is briefly discussed.
\end{abstract}

\maketitle

\section{Introduction}
The field of stochastic heat engines and refrigerators has seen intense activity in recent years~\cite{sei08_epl,Rana2014,Abah2014,Verley2014,bechinger2012,Koski2014,Abah2016,Goold2019,roldan2016}, largely owing to their potential applicabilities, especially in the health industry~\cite{vyas2014}. These engines are microscopic counterparts of heat engines that are commonly encountered in daily life. The bane in this microscopic world is the presence of appreciable amount of thermal fluctuations  significantly affecting the working of these engines~\cite{rit05_pt,rit06_jpcm}. The challenge lies in constructing engines at these scales that are comparable in efficiency to the molecular motors at work within our body cells~\cite{Astumian2001,Astumian2002,Astumian2010,Ait-Haddou2003}.

Presence of thermal fluctuations induces a stochastic time evolution of dynamical variables characterising the stochastic heat engines. Stochastic processes have always constituted a very active area of research in the domain of statistical physics. Over the years, a variety of model stochastic processes has been explored in the literature, the genesis of all of which lies in the paradigmatic Brownian motion. A satisfactory microscopic understanding of the Brownian motion was put forward by Einstein in one of his annus mirabilis papers~\cite{zwanzig2001nonequilibrium,balakrishnan2008elements}. Subsequently, the basic set-up of the Brownian motion has been generalized in many different directions of practical relevance, e.g., the Ornstein-Uhlenbeck process~\cite{uhlenbeck1930theory}, the Kramers problem~\cite{kramers1940brownian}, the phenomenological stochastic model for lineshapes in spectroscopy~\cite{kubo1969stochastic}, and many others. 

In recent times, stochastic resetting of the Brownian motion has emerged as an active area of study of stochastic processes; for a review, see \cite{r1,gupta2022stochastic,nagar2023stochastic,pal2022inspection}. Here, the usual Brownian dynamics is interspersed with an instantaneous reset of its position to its initial location at random times~\cite{evans2011diffusion}. Even this apparently simple modification of the dynamics has a dramatic and non-trivial effect on the emergent properties of the system. Indeed, in absence of resetting, the motion is unbounded and has a spatial probability distribution that is a Gaussian with a time-dependent width, resulting in the mean squared displacement (MSD) of the particle from its initial location increasing forever linearly with time. Introducing resetting into the dynamics leads to an effective confinement of the Brownian particle in space, in the sense that the dynamics at long times relaxes to a stationary state characterized by a time-independent spatial probability distribution. Consequently, the MSD of the particle does not increase forever as a function of time but instead saturates at long times to a time-independent value~\cite{r2}. Remarkably, the mentioned confinement takes place despite the fact that there are no actual physical boundaries in space confining the particle. It was soon realized that the potential of such a dynamical scenario in generating non-trivial static and dynamic properties may be explored in more general set-ups, namely, those that involve not one but many particles interacting with one another, and also the system of study could be undergoing bare evolution in absence of resetting according to any predefined dynamics of relevance, and not necessarily Brownian dynamics. Indeed, exploration of interesting and sometimes intriguing consequences of resetting has been pursued in a variety of diverse dynamical scenarios. A panorama of the applications of stochastic resetting in classical systems may be witnessed in the representative references, Refs.~\cite{Durang2014,kusmierz2014first,Gupta2014,Nagar2016,Pal2016a,Eule2016,pal2017first,Roldan2017,Shkilev2017,
Masoliver2019,Masoliver2019a,campos2006transport,Pal2019,Hollander2019,Magoni2020,belan2018restart,Bruyne2020,Karthika2020,Sadekar2020,Singh2020a,Bressloff2020,Grange2020,Tucci2020,Plata2020,MercadoVasquez2021,Dahlenburg2021,Singh2021,Wang2021,Singh2022,
Sarkar2022,Sarkar2022a,Giuggioli:2022}

In the above backdrop, we address here an hitherto unexplored issue of relevance: how does a stochastic heat engine perform when subject to stochastic resetting? Can one make an engine more efficient through introduction of stochastic resetting or does the latter prove to be detrimental and a nuisance? Since effects of stochastic resetting were first unveiled in the context of Brownian motion, it behooves us to choose the working system for our engine to be a Brownian particle moving in one dimension in presence of a harmonic potential and in contact with two heat baths at a colder and a hotter temperature. The expansion and the compression step of the engine are implemented through a time-dependent manipulation of the stiffness coefficient of the harmonic potential \cite{Schmiedl_2007,bechinger_2012,Saha2019,lahiri2020}. The dynamics involves the Brownian particle undergoing overdamped Langevin dynamics in presence of the harmonic potential with a time-varying stiffness coefficient and in contact with either the hotter or the colder heat bath; the dynamics is interrupted at random times with a stochastic resetting of the particle. The time intervals $\tau$ between successive resetting are random variables that are taken to be sampled independently from an exponential distribution $p(\tau)=re^{-r\tau}$, where $r \ge 0$ is the resetting rate, i.e., the probability per unit time for a resetting to take place. 
We primarily consider the case where the resetting takes place to the minimum of the harmonic potential, and later briefly explore the case where the resetting is to locations other than the minimum.  

Our principal findings are as follows: we find that indeed resetting does play a constructive role in rendering a stochastic heat engine more efficient. The efficiency increases monotonically as a function of $r$ for the case of resetting to the minimum of the harmonic potential. Interestingly, the work output from the engine is found to be a non-monotonic function of the resetting rate $r$ for small values of the  stiffness coefficient of the harmonic potential.
On the other hand, when the resetting takes place to a location other than the potential minimum, the efficiency still increases with $r$, but which for a fixed $r$ decreases with the distance of the reset point from the minimum of the harmonic potential. 
 Along the way of pursuing our analysis, we discover and clarify crucial technical issues related to correct thermodynamic interpretation of physical quantities  involved in characterizing the energetics and the efficiency of an engine. Thus, our contributions are two-fold: on one hand, to study the working of a heat engine subject to stochastic resetting, and on the other, to lay down the correct framework for its theoretical analysis. 

The paper is laid out as follows: In Section~\ref{sec:model}, we discuss our model of stochastic heat engine and present some preliminary analytical results that form the core of the analysis of the thermodynamics of the heat engine, which we take up in Section~\ref{sec:thermodynamics}. Results of our analysis are presented in Section~\ref{sec:results}, while the paper ends with conclusions in Section~\ref{sec:conclusions}.

\section{Model}
\label{sec:model}
As mentioned in the introduction, our engine employs as a system a single Brownian particle in one dimension. Let the variable $x(t)$ denote its location at time $t$. The particle is undergoing overdamped Langevin dynamics in presence of trapping due to a harmonic potential $V(x,t)$ with a periodically-varying stiffness parameter $k(t)$,
 \begin{align}
V(x,t)=\frac{1}{2}k(t)x^2,
 \end{align}
 and in presence of a heat bath in equilibrium at temperature $T$. The dynamics is repeatedly interrupted at exponentially-distributed random time intervals by a stochastic reset of the current location of the particle to a given location $x_\mathrm{r}$. Specifically, the dynamics of the location $x(t)$ of the particle, while starting at time $t=0$ from an initial location $x_0\equiv x(t=0)$, involves the following: in the small interval $[t,t+\mathrm{d}t]$, with a probability of $1-r \mathrm{d}t$ the particle undergoes an evolution following the familiar Langevin equation:
\begin{align}
    \gamma\frac{\mathrm{d}x}{\mathrm{d}t}= -k(t)x + \xi(t),
    \label{eq:eom}
\end{align}
or resets to the location $x_\mathrm{r}$ with the complementary probability $r\mathrm{d}t$. Here,  $\gamma>0$ is the damping coefficient, and  $\xi(t)$ denotes a Gaussian-distributed white noise with zero mean, $\langle \xi(t) \rangle=0$, where angular brackets denote averaging over noise realizations. The temporal correlations of the noise are given by $\langle \xi(t)\xi(t') \rangle = 2D\delta(t-t')$, with $D=\gamma k_B T$, and $k_B$ being the Boltzmann constant. The parameter $r\ge 0$ is the resetting rate, i.e., the probability per unit time for undergoing a reset.
Setting $r$ to zero reduces the dynamics to the bare dynamics~\eqref{eq:eom} without resets. Using the definition of $r$, it is easily checked that the random interval $\tau$ between two successive resets is distributed as an exponential: $p(\tau)=re^{-r\tau}$. The dynamics of the location of the particle may be summarized as follows:
\begin{align}
x(t+\mathrm{d}t)=\left\{ 
\begin{array}{ll}
x_\mathrm{r}~\mbox{with~probability~$r\mathrm{d}t$}, \\
               x(t)(1-\frac{k(t)}{\gamma}\mathrm{d}t)+\frac{1}{\gamma}\int_t^{t+\mathrm{d}t}\mathrm{d}t'~\xi(t')\\~\mbox{with probability~$1-r\mathrm{d}t$}.
               \end{array}
        \right. 
        \label{eq:updations of x}
\end{align}
As has been emphasized, resetting serves as a protocol for nonequilibrium drive in the sense that any steady state it induces is a genuine nonequilibrium steady state~\cite{r1}, and that the total entropy production rate under resetting is positive \cite{fuchs2016stochastic,gupta2022stochastic}.
We choose $x_0=0$, namely, the initial location of the particle coincides with the minimum of the potential $V(x,t)$.

 Our heat engine works in a Stirling cycle \cite{zem,bechinger_2012,Saha2019,lah2021,lahiri2020}. Starting at time $t=0$ with the Brownian particle at location $x_0=0$, the engine is run for several cycles, with each cycle being of total duration $\mathcal{T}$ that comprises the four steps enumerated below. 

\noindent (i) \textbf{Isothermal expansion:} During this step, the Brownian particle undergoes evolution~\eqref{eq:eom} interspersed with stochastic resetting for a total duration of time $\mathcal{T}/2$ during which the heat bath in contact is in equilibrium at a fixed temperature $T_H$, while the stiffness parameter $k$ decreases linearly with time as follows:
 \begin{align}
k(t)=k_\mathrm{exp}(t) &=k_0(1-t/\mathcal{T}); ~~t\in [0,\mathcal{T}/2],~k_0>0.
\label{eq:k-expansion}
\end{align} 
From the second cycle onwards, the time $t$ (but \textit{not} the position $x$) is reset to zero at the beginning of each cycle, so as to restrict its value for the isothermal expansion step to lie within the range $[0,\mathcal{T}/2]$. 
It is evident that the aforementioned decrease in the value of the stiffness coefficient (from $k_0$ to $k_0/2$ in time $\mathcal{T}/2$) allows the particle the possibility to travel further with respect to the potential minimum, thereby mimicking an expansion process \cite{sei08_epl,Saha2019,lahiri2020}. 

\noindent (ii) \textbf{Isochoric cooling:} During this step, the heat bath at temperature $T_H$ is replaced with one in equilibrium at a lower temperature $T_C<T_H$, with the stiffness parameter held fixed at the value $k_0/2$. In our study, we assume this step to be instantaneous, and so the particle location does not change during this step. 

\noindent (iii) \textbf{Isothermal compression:} During this step that lasts during the interval $[\mathcal{T}/2,\mathcal{T}]$, the dynamics of evolution~\eqref{eq:eom} interspersed with stochastic resetting proceeds in presence of the heat bath in equilibrium at temperature $T_C$ and with the stiffness parameter $k(t)$
that increases linearly with time (from $k_0/2$ to $k_0$ in time $\mathcal{T}/2$) and consequently constrains the motion of the particle:
\begin{align}
k(t)=k_\mathrm{com}(t)= k_0 t/\mathcal{T};~~t\in[\mathcal{T}/2,\mathcal{T}].
\label{eq:k-compression}
\end{align} 
As mentioned following Eq.~\eqref{eq:k-expansion}, we reset time to zero at the beginning of every cycle, from second cycle onwards; this restricts its value for the isothermal compression step to always lie within the range $[\mathcal{T}/2,\mathcal{T}]$.

\noindent (iv) \textbf{Isochoric heating:} During this step, which takes place instantaneously in time, the heat bath at temperature $T_C$ is replaced with one in equilibrium at temperature $T_H$. The stiffness coefficient is held fixed at the value $k_0$. The location of the particle does not undergo any change during this step.

In the aforementioned protocol, it is evident that work is done on the system during the compression step while increasing the value of the stiffness parameter. On the other hand, work is done by the system (i.e., energy is extracted from the system) during the expansion step. The net effect of the above steps is extraction of an average work (averaged over an ensemble of dynamical realizations, all involving running the dynamics with the same protocol for change of $k(t)$ applied periodically with period $\mathcal{T}$), and it is this feature that endows the system with the possibility to be used as an engine.
It is evident that in a given dynamical realization, we have $x_0 \equiv x(t=0)=0$ for the first cycle, but which for subsequent cycles has the value identical to the value of the particle position reached at the end of the previous cycle. This information may be encoded by defining a probability distribution $P_i(x_0)$ for the location of the Brownian particle at time $t=0$. 
For the first cycle, we have $P_i(x_0)=\delta(x_0)$. For the subsequent cycles, the distribution would have a finite width. For instance, at the end of the first cycle, one records the values of $x$ for different realizations. These values, which generally vary from one realization to another, generate the distribution $P_i(x_0)\ne \delta(x_0)$ for the second cycle. 

Let us note that there are two sources of stochasticity in the working of our heat engine. One is the presence of the heat bath that induces noise into the dynamics of the particle, while the other is the resetting of the location of the Brownian particle taking place at random times.

 Experimentally, stochastic heat engines (without resetting) have been frequently prepared by generating harmonic traps of temporally-modulated stiffness parameter by means of optical tweezers with tunable intensity~ \cite{bechinger2012,Krishnamurthy2016,Ganapathy2021}. The resetting process can be effected by a separate laser of high intensity \cite{pal2020experimental}, which forces the particle to quickly (compared to its time of relaxation to equilibrium in the harmonic trap in the absence of resets) locate to the location $x_\mathrm{r}$. 

\paragraph*{Variance in absence of resetting:}
It befits for later calculation to discuss here some salient features of the dynamics of the Brownian particle in absence of resetting and for time-independent stiffness coefficient $k(t)=k$. In this case, the particle dynamics, given by Eq.~\eqref{eq:eom}, is easily solved to obtain 
\begin{align}
    x(t)=\frac{1}{\gamma}\int_0^t \mathrm{d}t'~e^{-(k/\gamma)(t-t')}\xi(t'),
\end{align}
where we have taken the particle to have the initial position $x_0=0$. Noting that $\langle x(t)\rangle=0$, the mean squared displacement (MSD) is obtained as
\begin{align}
\sigma_\mathrm{nr}(t) &\equiv \langle x^2(t)\rangle - \langle x(t)\rangle^2 \nonumber\\
&=\langle x^2(t)\rangle \nonumber \\
&= \frac{k_{B}T}{k}(1-e^{-2kt/\gamma}),
\label{eq:MSD-noreset}
\end{align}
where the subscript ``$\mathrm{nr}$'' implies that the no-reset case is being considered.

We now include the effects of resetting, still with a time-independent $k$. To address the situation at hand, we invoke the renewal equation approach for the conditional probability density $P(x,t|x_0,0)$ to find the particle at location $x$ at time $t$, conditioned on having been at $x_0$ at time $t=0$. For the case $x_0=0=x_\mathrm{r}$, the renewal equation reads~\cite{evans2011diffusion,evans2020stochastic,gupta2022stochastic}
\begin{align}
&P_\mathrm{r}(x,t|0,0)=P_\mathrm{nr}(x,t|0,0)e^{-rt}\nonumber \\
&+r\int_0^ t \mathrm{d}\tau~e^{-r(t-\tau)}P_\mathrm{nr}(x,t|0,\tau), 
\label{eq:renewal-equation}
\end{align}
where the subscript ``$\mathrm{r}$'' is to emphasize that the corresponding result holds in presence of resetting, $\tau$ is the time instant at which the \emph{last} reset has taken place with respect to the time instant $t$ of observation, and $P_\mathrm{nr}(x,t|x_0,0)$ is the conditional probability density in absence of resetting. Equation~\eqref{eq:renewal-equation} has the following interpretation. To be at location $x$ at time $t>0$, the particle must either (i) not have undergone a single reset since the initial time instant, or, (ii) that the last reset happened during the time interval $[\tau-\mathrm{d}\tau,\tau];~\tau \in [0,t]$, and thereafter the particle has undergone free evolution up to time $t$. For case (i), the probability for no reset to take place during time duration $t$ is $\int_t^\infty\mathrm{d}\tau~p(\tau)=e^{-rt}$. For case (ii), the probability for the last reset to have happened during the time interval $[\tau-\mathrm{d}\tau,\tau]$ is $r\mathrm{d}\tau~e^{-r(t-\tau)}$. Using these results, one obtains the first and the second term on the right-hand side of Eq.~\eqref{eq:renewal-equation}. 

Given that one has $x_0=0=x_\mathrm{r}$, it follows from Eqs.~\eqref{eq:eom} and~\eqref{eq:updations of x} that $\langle x(t)\rangle=0$ even in the presence of resetting. Using 
\begin{align}
  \int_{-\infty}^\infty \mathrm{d}x~x^2 P_\mathrm{nr}(x,t|0,0) &=  \sigma_\mathrm{nr}(t),
\end{align}
Eq.~\eqref{eq:renewal-equation} yields 
\begin{align}
\sigma_\mathrm{r}(t) &=e^{-rt}\sigma_\mathrm{nr}(t)+r\int_{0}^{t}\mathrm{d}\tau~e^{-r(t-\tau)}\sigma_\mathrm{nr}(t-\tau)\nonumber \\
&= \frac{2\left(1-e^{-rt-2kt/\gamma}\right)k_B T}{2k+\gamma r}.
\label{eq:analytical_variance}
\end{align} 
It can be readily verified that on setting $r=0$, we correctly recover Eq.~\eqref{eq:MSD-noreset}.

\section{Thermodynamics of the heat engine}
\label{sec:thermodynamics}
In this section, we take up the main objective of this work, namely, analysing the thermodynamics of our stochastic heat engine described in the preceding section. To this end, let us first define the relevant thermodynamic quantities of the engine in conformity with the prescription laid down in stochastic thermodynamics~\cite{sek98,sekimoto}, which is the branch of thermodynamics that generalizes the thermodynamic laws to take into account stochasticity or fluctuations in the dynamics of the underlying system; for a review, see Refs.~\cite{sei12_rpp,peliti2021stochastic}. The quantities of interest are (a) work output and power, (b) heat absorbed, and (c) efficiency. 
We choose the convention that the work \textit{done by} and the heat \textit{absorbed by} the system are positive, while the work done on the system and heat dissipated by the system are negative.
Let us now define these quantities in turn.

\noindent (a) \textbf{Work output:} The stochastic work obtained as an output from our engine may be defined as follows. In a small time $\Delta t$, the Brownian particle resets with probability $r\Delta t$ resulting in an energy change of amount $V(x_\mathrm{r},t)-V(x,t)$, while it does not reset with the complementary probability $(1-r\Delta t)$ resulting in an amount of work $-(\partial V/\partial t)(t)~\Delta t$ done by the system~\cite{sek98,sekimoto}. Then, if $W_\mathrm{tot}(t)$ is the amount of work done \textit{by the system} for duration $t$, we have
\begin{align}
    W_\mathrm{tot}(t+\Delta t) = & W_\mathrm{tot}(t)+r\Delta t [V(x,t)-V(x_\mathrm{r},t)] \nonumber\\
     & +(1-r\Delta t) \left(-\frac{\partial V}{\partial t}(t)~\Delta t\right),
\end{align}
yielding in the limit $\Delta t \to 0$ that
\begin{equation}
    \frac{\mathrm{d}W_\mathrm{tot}(t)}{\mathrm{d}t}=r[V(x,t)-V(x_\mathrm{r},t)]-\frac{\partial V}{\partial t}(t).
\end{equation}
On integrating the above equation with respect to time, we get the total work done by the system between times $t_i$ and $t_f$, with $t_f-t_i=t$, given by 
\begin{align}
     W_\mathrm{tot}(t) &= -\int_{t_i}^{t_f} \mathrm{d}t~\frac{\partial V}{\partial t}(t) + r\int_{t_i}^{t_f} \mathrm{d}t~[V(x,t)-V(x_\mathrm{r},t)],
    \label{eq:work}
\end{align} 
with
\begin{align}
 \frac{\partial V}{\partial t}(t)=\frac{1}{2}\dot{k}(t)x^2(t), 
 \label{eq:dVdt}
\end{align}
and the dot denoting derivative with respect to time.

It may be noted that we need to use the definition in Eq.~\eqref{eq:work} with caution. An example of a similar process in a macroscopic system may help in further clarification. Consider a cylinder one of whose walls is a movable piston, and which contains an ideal gas of \textit{charged} particles. When the piston is pushed inwards so as to compress the gas, work is done on  the system, while work is done by the system when the piston is pulled outwards~\cite{zem}. These pull-push moves constitute the usual protocol for operating an engine. Additionally, in order to implement resetting events, one can switch on a strong electric field at random times, such that in its presence, the particles are forced to collect near one of the walls. Every time such a reset event takes place, the gas particles colliding with the piston suddenly undergo a change in configuration (positions and momenta). This change is brought about by the electric field  and not by the motion of the piston, while it is the manipulation of the latter that constitutes the protocol for operating the engine. Therefore, the work done by the field in carrying out the reset is to be excluded in computing the work output from the engine. Nevertheless, the work output over a period of time will have an implicit and essential contribution from the fact that the system it is acting upon undergoes sudden changes in configuration arising from resetting at random times. 

With the above example in mind, we return to our microscopic system and in particular to Eq.~\eqref{eq:work}. We note from the equation that there are actually two sources of work done. The work that is done as part of the engine protocol is represented by the first term on the right hand side of the equation. The second term is due to the phenomenon of resetting. As discussed above, in computing the work extracted using the engine protocol only, we should exclude the second term. We then obtain the \textit{output work} $W_\mathrm{out}(t)$ from the engine as given by~\cite{Pal2020}
\begin{align}
    W_\mathrm{out}(t) &= -\int_{t_i}^{t_f} \mathrm{d}t~\frac{\partial V}{\partial t}(t).
    \label{eq:Wout}
\end{align}

\noindent \textbf{Power:} The power output of the engine in a cycle of its run is given by the average work extracted in the cycle divided by the time period of the cycle and is thus equal to the quantity $\langle W_\mathrm{out}(\mathcal{T})\rangle/\mathcal{T}$. Here and in the following, the angular brackets denote as usual the process of ensemble averaging of the otherwise stochastic quantities. 

\noindent (b) \textbf{Heat absorbed:} The internal energy of the overdamped Brownian particle is given by its potential energy $V(x,t)$. The change in the internal energy is accordingly defined for the evolution from time $t_i$ to time $t_f$ as 
\begin{align}
\Delta E(t) = V(x_f,t_f)-V(x_i,t_i),
\label{eq:internal-energy}
\end{align}
with $x_i=x(t_i)$ and $x_f=x(t_f)$. The average heat absorbed by the particle from the surrounding heat bath can then be obtained by using the first law of thermodynamics, as 
\begin{align}
\langle Q(t)\rangle = \langle W_\mathrm{out}(t)\rangle + \langle\Delta E(t)\rangle.
\label{eq:first-law}
\end{align} Computing these quantities for the expansion step of the engine yields the average heat $\langle Q_H(\mathcal{T}/2)\rangle$ absorbed during the expansion step. 

\noindent (c) \textbf{Efficiency:} The efficiency of the engine is defined in the usual manner, namely, that it is the average work $\langle W_\mathrm{out}(\mathcal{T})\rangle$ extracted per cycle, divided by the average heat $\langle Q_H(\mathcal{T}/2)\rangle$ absorbed per cycle during the expansion step of the engine:
\begin{align}
    \eta \equiv \frac{\langle W_\mathrm{out}(\mathcal{T})\rangle}{\langle Q_H(\mathcal{T}/2)\rangle}.
    \label{eq:eta}
\end{align}
One may compute the efficiency after the engine has been run for an integer number of cycles denoted by $p$.
In particular, one is interested in the efficiency in the time-periodic steady state (TPSS) of the engine, achieved in the limit $p \to \infty$, and which may be implemented in practice in the following way. Starting with $x_0=0$, a dynamical realization corresponds to running the engine for a very large number of cycles and then one computes the quantities $W_\mathrm{out}(\mathcal{T})$ and $Q_H(\mathcal{T}/2)$ for the last cycle. The process is repeated over and over again, and this allows to obtain the aforementioned quantities averaged over an ensemble of dynamical realizations, thereby yielding the desired efficiency $\eta$ in the TPSS.

We now turn to a discussion of the behavior of our model engine.

\section{Results and discussions}
\label{sec:results}

\subsection{Calculation of variance}
\label{sec:Variance}

In this case, throughout the engine cycle, the reset location $x_\mathrm{r}$ has the value $x_\mathrm{r}=0$, which is also the minimum of the harmonic potential. As already discussed earlier, we have $x_0=0$ for the first cycle, while for subsequent cycles, $x_0$ in a given dynamical realization of the engine is the location reached at the end of the previous cycle in the same dynamical realization. 

In order to obtain the efficiency, we start with the renewal equation~\eqref{eq:renewal-equation}, which in the present case reads for the expansion (exp) and the compression (com) step in a given realization as
\begin{align}
&\mathcal{P}^\mathrm{exp}_\mathrm{r}(x,t|x_0,0)=\mathcal{P}^\mathrm{exp}_\mathrm{nr}(x,t|x_0,0)e^{-rt}\nonumber \\
&+r\int_0^ t \mathrm{d}\tau~e^{-r(t-\tau)}\mathcal{P}^\mathrm{exp}_\mathrm{nr}(x,t|0,\tau);~0\le t \le \mathcal{T}/2, \nonumber \\
\label{eq:renewal-equation-1} \\
&\mathcal{P}^\mathrm{com}_\mathrm{r}(x,t|x(\mathcal{T}^+/2),\mathcal{T}^+/2) \nonumber\\
&=\mathcal{P}^\mathrm{com}_\mathrm{nr}(x,t|x(\mathcal{T}^+/2),\mathcal{T}^+/2)e^{-r(t-\mathcal{T}^+/2)}\nonumber\\
&+r\int_{\mathcal{T}/2}^ t \mathrm{d}\tau~e^{-r(t-\tau)}\mathcal{P}^\mathrm{com}_\mathrm{nr}(x,t|0,\tau);~\mathcal{T}/2 \le t \le \mathcal{T}.\nonumber 
\end{align}
Here, the conditional probability density $P_\mathrm{r}^{\mathrm{exp}}(x,t|x_0,0)$ gives the probability of the particle to be at position $x$ at time $t$, given that it was at position $x_0$ at time $t=0$, during the expansion step (denoted by the superscript ``$\mathrm{exp}$'') and in presence of resetting (denoted by the subscript ``$\mathrm{r}$''). All the other conditional probabilities are similarly defined. The symbols $\mathcal{T}^-$ and $\mathcal{T}^+$ denote the time instants just before and just after the time instant $t=\mathcal{T}$.

We further stipulate that
\begin{align}
&\mathcal{P}^\mathrm{com}_\mathrm{nr}(x,\mathcal{T}^+/2|x(\mathcal{T^-}/2),\mathcal{T^-}/2)=\delta(x-x(\mathcal{T^-}/2)),
\label{eq:conditions} 
\end{align}
which implies that the initial value of the particle location for the compression step is that reached at the end of the expansion step corresponding to the same realization of the engine cycle. Similarly, due to our convention regarding $x_0$ mentioned in the first paragraph of this subsection, we also stipulate that
\begin{align}
    \mathcal{P}_\mathrm{nr}^\mathrm{exp}(x_0,\mathcal{T}^+|x(\mathcal{T}^-),\mathcal{T}^-) &= \delta(x_0-x(\mathcal{T}^-)).
    \label{eq:conditions end-point}
\end{align}

Let us now make some important remarks. In a given dynamical realization, the particle in the first cycle starts from $x_0=0$, which is the minimum of the harmonic potential that has the $x\to -x$ symmetry, and resets repeatedly to $x_\mathrm{r}=0$. Consequently, at any time  $t$ during the first cycle, the ensemble average is $\langle x(t)\rangle=0$ and the random variable $x(t)$ has a distribution that is symmetric about zero. In the second cycle, the particle starts from $x(\mathcal{T}^-)$ and resets repeatedly to $x_\mathrm{r}=0$. It is evident from the above that we have $\langle x(\mathcal{T}^-)\rangle=0$ and the distribution of $x(\mathcal{T}^-)$ is symmetric about zero. Consequently, at any time $t$ during the second cycle, one has the ensemble average $\langle x(t)\rangle=0$ and the distribution of $x(t)$ is symmetric about zero. Arguing in this manner, it readily follows that similar features hold for every cycle. Given these facts, three different MSDs can be defined as follows (which appear later in Eq. \eqref{eq:renewal-equation-2}):
\begin{align}
    &\sigma^\mathrm{exp}_\mathrm{r}(t)= \int_{-\infty}^{+\infty} \mathrm{d}x_0 \int_{-\infty}^{+\infty}\mathrm{d}x~ x^2  P^\mathrm{exp}_\mathrm{r}(x,t|x_0,0)P_i(x_0), \nonumber\\
    &\sigma^\mathrm{exp}_\mathrm{nr}(t) = \int_{-\infty}^{+\infty} \mathrm{d}x_0 \int_{-\infty}^{+\infty}\mathrm{d}x~ x^2  P^\mathrm{exp}_\mathrm{nr}(x,t|x_0,0)P_i(x_0), \nonumber\\
    &\tilde\sigma^\mathrm{exp}_\mathrm{nr}(t-\tau) = \int_{-\infty}^{+\infty} \mathrm{d}x ~ x^2  P^\mathrm{exp}_\mathrm{nr}(x,t|0,\tau). \nonumber\\
    \label{eq:SigmaDefinitions}
\end{align}
The third definition is for the variance obtained after the particle has been reset to $x=0$ at time $t=\tau$ and is thereafter allowed to evolve till time $t$.
Similar expressions can be readily written down for $\sigma^\mathrm{com}_\mathrm{r}(t)$, $\sigma^\mathrm{com}_\mathrm{nr}(t)$ and $\tilde\sigma^\mathrm{com}_\mathrm{nr}(t)$, as
\begin{align}
    &\sigma^\mathrm{com}_\mathrm{r}(t)= \int_{-\infty}^{+\infty} \mathrm{d}x(\mathcal{T}^+/2) \int_{-\infty}^{+\infty}\mathrm{d}x~ x^2  \nonumber \\
    &\hspace{0.5cm}\times P^\mathrm{com}_\mathrm{r}(x,t|x(\mathcal{T}^+/2),\mathcal{T}^+/2)P(x(\mathcal{T}^+/2)), \nonumber\\
    &\sigma^\mathrm{com}_\mathrm{nr}(t) = \int_{-\infty}^{+\infty} \mathrm{d}x(\mathcal{T}^+/2) \int_{-\infty}^{+\infty}\mathrm{d}x~ x^2  \nonumber \\
    &\hspace{0.5cm}\times P^\mathrm{com}_\mathrm{nr}(x,t|x(\mathcal{T}^+/2),\mathcal{T}^+/2)P(x(\mathcal{T}^+/2)), \nonumber\\
    &\tilde\sigma^\mathrm{com}_\mathrm{nr}(t-\tau) = \int_{-\infty}^{+\infty} \mathrm{d}x ~ x^2  P^\mathrm{com}_\mathrm{nr}(x,t|0,\tau), \nonumber\\
    \label{eq:SigmaDefinitions1}
\end{align}
where $P(x(\mathcal{T}^+/2))$ is the distribution of the random variable $x(\mathcal{T}^+/2)$, which, as argued above, is symmetric about zero.

From Eq.~\eqref{eq:renewal-equation-1} we then get, on using the normalization of the distributions $P_i(x_0)$ and $P(x(\mathcal{T}^+/2))$, 
\begin{align}
\sigma^\mathrm{exp}_\mathrm{r}(t) &=\sigma^\mathrm{exp}_\mathrm{nr}(t)e^{-rt}\nonumber \\
&+r\int_0^ t \mathrm{d}\tau~e^{-r(t-\tau)}\tilde\sigma^\mathrm{exp}_\mathrm{nr}(t-\tau);~~0 \le t \le \mathcal{T}/2, \nonumber \\
\label{eq:renewal-equation-2} \\
\sigma^\mathrm{com}_\mathrm{r}(t)&=\sigma^\mathrm{com}_\mathrm{nr}(t)e^{-r(t-\mathcal{T}/2)}\nonumber \\
&+r\int_{\mathcal{T}/2}^t \mathrm{d}\tau~e^{-r(t-\tau)}\tilde\sigma^\mathrm{com}_\mathrm{nr}(t-\tau);~~\mathcal{T}/2 \le t \le \mathcal{T}, \nonumber 
\end{align}
with 
\begin{align}
\sigma^\mathrm{com}_\mathrm{r}(\mathcal{T^+}/2) = \sigma^\mathrm{com}_\mathrm{nr}(\mathcal{T^+}/2)=\sigma_\mathrm{r}^\mathrm{exp}(\mathcal{T^-}/2).
\label{eq:VarianceMatching}
\end{align}
Since $\sigma^\mathrm{exp}_\mathrm{r}(\mathcal{T^-}/2)$ is the variance reached at the end of the expansion step, it is also the initial variance for the compression step. This initial variance would be the initial condition to evaluate both $\sigma^\mathrm{com}_\mathrm{nr}(t)$ and $\sigma^\mathrm{com}_\mathrm{r}(t)$, for $t>\mathcal{T}/2$. This is because the former quantity is the variance in absence of resetting \textit{only in the compression step}, with the expansion step having already been performed in the \textit{presence} of resetting.

To proceed, we need the quantities $\sigma^\mathrm{exp}_\mathrm{nr}(t)$ and $\sigma^\mathrm{com}_\mathrm{nr}(t)$, with $t$ lying in the appropriate ranges mentioned in Eq.~\eqref{eq:renewal-equation-2}. To this end, we first solve Eq.~\eqref{eq:eom} with the corresponding time-dependences given by Eqs.~\eqref{eq:k-expansion} and~\eqref{eq:k-compression}. The  formal solution reads 
\begin{align}
    x(t) &= x_0e^{-I(t,0)} + \frac{e^{-I(t,0)}}{\gamma}\int_{0}^t \mathrm{d}t'~ \xi(t')e^{I(t',0)},
\end{align}
with $I(t,t_0) \equiv \int_{t_0}^t \mathrm{d}t~k(t)/\gamma$. 
The MSD is immediately obtained as
\begin{align}
    \sigma_\mathrm{nr}(t) &= \sigma_\mathrm{nr}(0)e^{-2I(t,0)} \nonumber\\
     &+ \frac{2D}{\gamma^2}e^{-2I(t,0)}\int_{0}^t \mathrm{d}t'~e^{2I(t',0)}.
\label{eq:GeneralExpression}
\end{align}
For instance, for the expansion step of the first cycle for which we have $x_0=0$, we obtain
\begin{align}
    &\sigma_\mathrm{nr}^\mathrm {exp}(t) = D_H\sqrt{\frac{\pi\mathcal{T}}{k_0\gamma^3}} \exp \left[-\frac{2k_0 t}{\gamma}\left(1-\frac{t}{2\mathcal{T}}\right) + \eta^2(0)\right] \nonumber\\
    &\vspace{0.5cm}\times (\mathrm{erf}[\eta(0)]-\mathrm{erf}[\eta(t)]),
\end{align}
where we have $D_H=\gamma k_B T_H$, we have defined the function
\begin{align}
    \eta(s) \equiv \sqrt{\frac{k_0}{\gamma\mathcal{T}}}~(\mathcal{T}-s),
\end{align}
and $\mathrm{erf}(x)=(2/\sqrt{\pi})\int_x^\infty \mathrm{d}y~e^{-y^2}$ is the error function. Similar expressions can be written for $\sigma^\mathrm{com}_\mathrm{nr}(t)$. 
Using Eq.~\eqref{eq:GeneralExpression} in Eq.~\eqref{eq:renewal-equation-2}, one arrives at formal analytical expressions for $\sigma_\mathrm{r}^\mathrm{exp}(t)$ and $\sigma_\mathrm{r}^\mathrm{com}(t)$ that can be evaluated numerically. 
For the subsequent cycles, we have $\sigma_\mathrm{nr}^\mathrm{exp}(\mathcal{T}^+) = \sigma_\mathrm{r}^\mathrm{com}(\mathcal{T}^-)$. In other words, the variance in position retains its continuity with time when the $(n-1)^\mathrm{th}$ cycle gives way to the $n^\mathrm{th}$ cycle.

\begin{figure}[h!]
\centering
\includegraphics[width=0.45\textwidth]{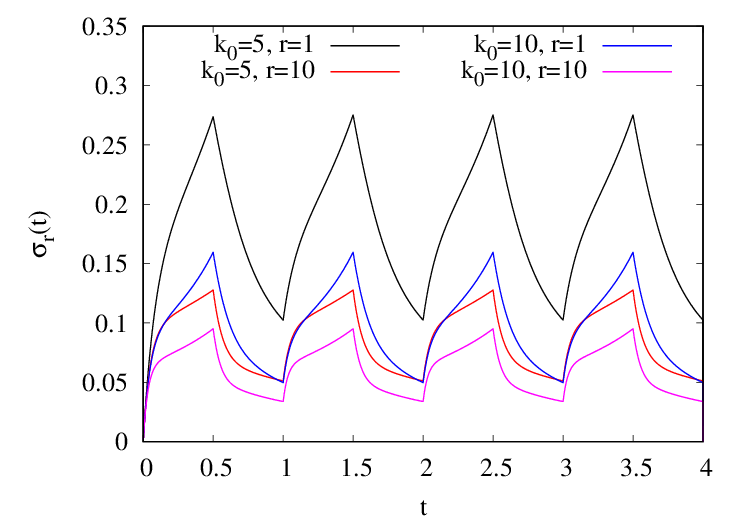}
    \caption{Plots for the quantity $\sigma_\mathrm{r}(t)$ (with $\sigma_\mathrm{r}(t)=\sigma_\mathrm{r}^\mathrm{exp}(t)$ and $\sigma_\mathrm{r}(t)=\sigma_\mathrm{r}^\mathrm{com}(t)$ in respectively the rising and the falling part of each curve), evaluated using Eq.~\eqref{eq:renewal-equation-2} and plotted as a function of time $t$ for different combinations of $k_0$ and $r$. One may observe the initial transient region and the subsequent settling into a time-periodic steady state (TPSS). Other parameters are $\mathcal{T}=1.0,~\gamma=1.0,~T_H=1.0,~T_C=0.5,~x_\mathrm{r}=0$.}
    \label{fig:TPSS_sigma}
\end{figure}

\begin{figure}[h!]
\centering\includegraphics[width=0.45\textwidth]{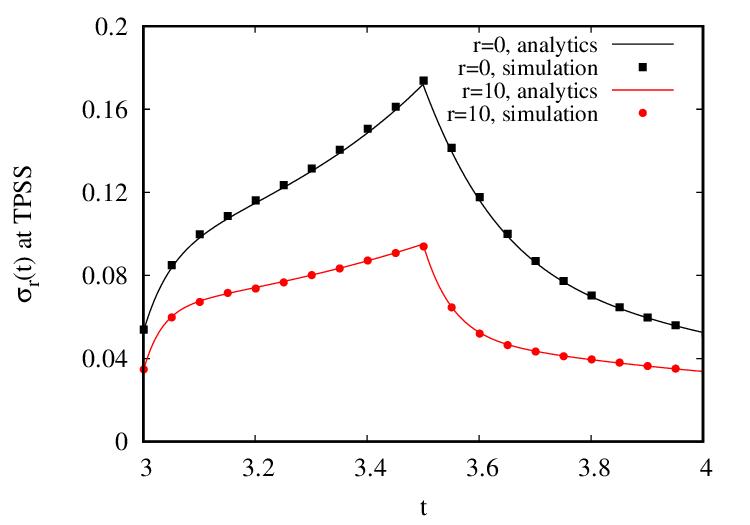}
    \caption{Plots for the quantity $\sigma_\mathrm{r}(t)$ (with $\sigma_\mathrm{r}(t)=\sigma_\mathrm{r}^\mathrm{exp}(t)$ and $\sigma_\mathrm{r}(t)=\sigma_\mathrm{r}^\mathrm{com}(t)$ in respectively the rising and the falling part of each curve) plotted as a function of time $t$ in the TPSS. The solid lines depict numerical results obtained using Eq.~\eqref{eq:renewal-equation-2}, while results from simulations of the dynamics of our engine are depicted by symbols. Here, $k_0=10$, the $r$-values are indicated in the figure, while the values of the other parameters are the same as in Fig.~\ref{fig:TPSS_sigma}.} \label{fig:comparison-sigma-TPSS}
\end{figure}

As remarked earlier, we are interested in the output of the engine in the TPSS, where the position distribution of the overdamped particle becomes periodic in time, with the period equaling $\mathcal{T}$. In discussing our results, we will always leave out the initial transients (where the distribution is yet to become periodic in time) and focus on the performance of the engine in its TPSS. The initial transient regime and the TPSS have been shown in Fig.~\ref{fig:TPSS_sigma} for the quantities $\sigma_\mathrm{r}^\mathrm{exp}(t)$ and $\sigma_\mathrm{r}^\mathrm{com}(t)$ and for various combinations of $k_0$ and $r$. The data are obtained from numerical evaluation of Eq.~\eqref{eq:renewal-equation-2}. It may be observed that the engine almost always enters into its TPSS right from the beginning of the second cycle. Nevertheless, in obtaining and discussing our results on the performance of the engine, we have always discarded the first three cycles and recorded our observations from the fourth cycle. The figure also shows that an increase in the value of $k_0$ at a fixed $r$ or an increase in the value of $r$ at a fixed $k_0$ leads to a decrease in the variance, in accordance with our expectations.

Numerical evaluation of $\sigma_\mathrm{r}^\mathrm{exp}(t)$ and $\sigma_\mathrm{r}^\mathrm{com}(t)$ 
 in the TPSS, by using Eq.~\eqref{eq:renewal-equation-2}, has been carried out and compared with the results obtained from our simulations of the engine dynamics in Fig.~\ref{fig:comparison-sigma-TPSS} for two different values of the resetting rate, namely $r=0$ and $r=10$. An excellent agreement between theory (solid lines) and simulations (symbols) is observed, thereby validating the accuracy of our simulations vis-\`{a}-vis our analytical results. 

\subsection{Variation of output work and efficiency with the resetting rate}
\label{sec:WorkAndEfficiency}

Using Eq.~\eqref{eq:dVdt} in Eq.~\eqref{eq:Wout}, one obtains the average output work as
\begin{align}
    \langle W_\mathrm{out}(\mathcal{T})\rangle =& -\frac{1}{2}\int_{0}^{\mathcal{T}/2} \mathrm{d}t~\dot k_\mathrm{exp}(t)\sigma_r^\mathrm{exp}(t)\nonumber\\
    &-\frac{1}{2}\int_{\mathcal{T}/2}^{\mathcal{T}} \mathrm{d}t~\dot k_\mathrm{com}(t)\sigma_r^\mathrm{com}(t).
    \label{eq:Wout_explicit}
\end{align}
The heat absorbed during the expansion step can be readily computed by using Eq.~\eqref{eq:first-law}, as
\begin{align}
    \langle Q_H(\mathcal{T}/2)\rangle_\mathrm{exp} &= \langle W_\mathrm{out}(\mathcal{T}/2)\rangle_\mathrm{exp} + \langle\Delta E(\mathcal{T}/2)\rangle_\mathrm{exp},
    \label{eq:QH}
\end{align}
where the subscripts outside the angular brackets imply that the corresponding quantities are to be calculated only for the expansion step. Here, on using Eq.~\eqref{eq:internal-energy}, we have
\begin{align}
    \langle\Delta E(\mathcal{T}/2)\rangle_\mathrm{exp} &= \frac{k_0\sigma_\mathrm{exp}(\mathcal{T}/2)}{4} - \frac{k_0\sigma_\mathrm{exp}(0)}{2}.
    \label{eq:}
\end{align}
The above expressions may be evaluated to obtain the functional dependency of the work output $\langle W_\mathrm{out}(\mathcal{T})\rangle$ and the efficiency $\eta$ on the resetting rate.

We begin our discussion of the results obtained by demonstrating that using $W_\mathrm{tot}(\mathcal{T})$ instead of $W_\mathrm{out}(\mathcal{T})$ while computing the engine efficiency (see Eqs.~\eqref{eq:work} and~\eqref{eq:Wout}) leads to unphysical results. We have 
\begin{align}
    &\langle W_\mathrm{tot}(\mathcal{T})\rangle = \langle W_\mathrm{out}(\mathcal{T})\rangle\nonumber \\
    &+ \frac{r}{2}\int_{0}^{\mathcal{T}/2} \mathrm{d}t~[k_\mathrm{exp}(t)\sigma_r^\mathrm{exp}(t) - k_0 \sigma_r^\mathrm{exp}(0)]\nonumber\\
    &+ \frac{r}{2}\int_{\mathcal{T}/2}^{\mathcal{T}} \mathrm{d}t~ [k_\mathrm{com}(t)\sigma_r^\mathrm{com}(t) - (k_0/2) \sigma_r^\mathrm{com}(\mathcal{T}/2)].
    \label{eq:wout-average}
\end{align}
The quantity $\langle W_\mathrm{tot}(\mathcal{T})\rangle$ is expected to increase as a function of $r$ due to the interplay between the two terms appearing on the right-hand side of Eq.~\eqref{eq:work}. Indeed, the second term on the right hand clearly shows an explicit dependence on $r$. The first term has an implicit dependence on $r$: the higher the resetting rate $r$, the lower is the possibility of the particle making large jumps in energy in small time steps (given that it becomes more difficult for the particle to access the steeper parts of the potential). This increases the contribution coming from the first term as well, leading to a larger $\langle W_\mathrm{out}(\mathcal{T})\rangle$ with higher value of $r$.

\begin{figure}[h!]
    \centering
    \includegraphics[width=0.45\textwidth]{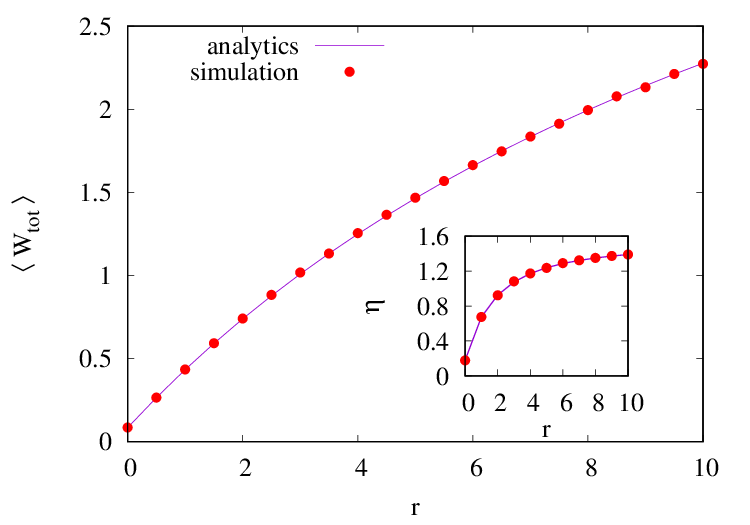}
    \caption{Numerical and simulation results for the variation of the output work with $r$, for $k_0=10$ and in the TPSS; the values of the other parameters are the same as in Fig.~\ref{fig:TPSS_sigma}. The numerical results correspond to Eq.~\eqref{eq:wout-average}. The output work is taken to be $\langle W_\mathrm{tot}(\mathcal{T})\rangle$.  Inset: Variation of engine efficiency $\eta$, Eq.~\eqref{eq:eta}, with $r$ and in the TPSS; in the equation, $W_\mathrm{out}$ has been replaced with $W_\mathrm{tot}$. The heat absorbed $\langle Q_H(\mathcal{T}/2)\rangle$ has been obtained from numerical evaluation of the expression in Eq.~\eqref{eq:QH}.}
    \label{fig:W_tot_r}
\end{figure}

In Fig.~\ref{fig:W_tot_r}, we plot the extracted work versus the rate of resetting, by (incorrectly) replacing $W_\mathrm{out}$ by $W_\mathrm{tot}$. In accordance with our aforementioned expectations, $\langle W_\mathrm{tot}(\mathcal{T})\rangle$ is observed to increase with the resetting rate. 
As may be observed from the inset, the corresponding efficiency also increases as the resetting rate is increased, and begins to saturate to a value \textit{higher than unity}, which we now show to be leading to mathematical inconsistency. We have
\begin{align}
    &1-\frac{|\langle Q_C\rangle|}{|\langle Q_H\rangle|} > 1 
    \hspace{0.3cm}\Rightarrow \hspace{0.3cm}\frac{|\langle Q_C\rangle|}{|\langle Q_H\rangle|} < 0,
    \label{eq:violation}
\end{align}
where $Q_C$ is the heat released into the cold bath in the compression step, while $Q_H \equiv Q_H(\mathcal{T}/2)$ is the heat absorbed from the hot bath in the expansion step. The relation \eqref{eq:violation} is clearly mathematically inconsistent.
Evidently, it was incorrect to naively assume $W_\mathrm{tot}$ to be the extracted work while computing the efficiency of the engine.

\begin{figure}[ht!]
    \centering
    \begin{subfigure}{0.45\textwidth}
    \includegraphics[width=\textwidth]{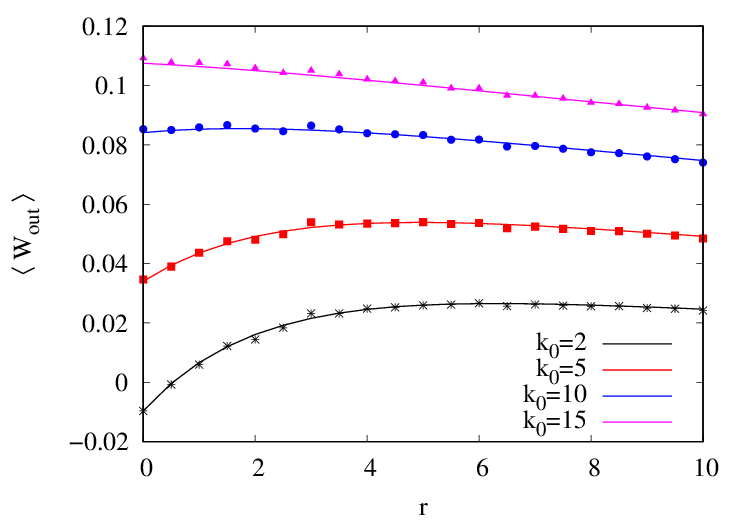}
    \caption{}
    \end{subfigure}
    \begin{subfigure}{0.45\textwidth}
    \includegraphics[width=\textwidth]{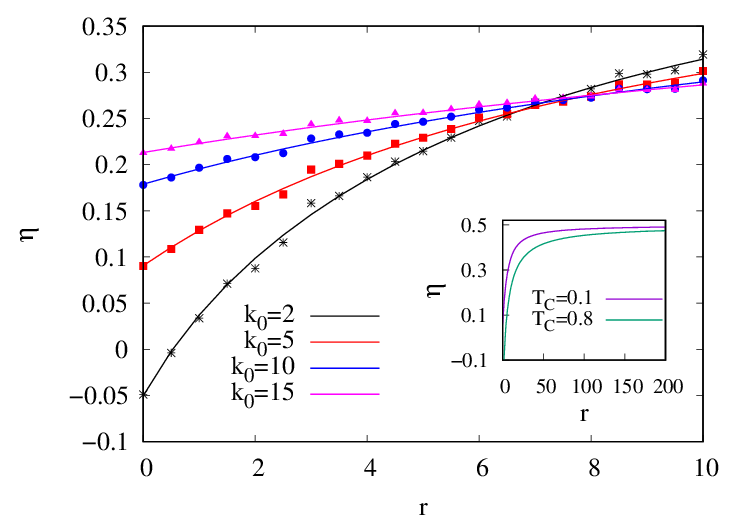}
    \caption{}
    \end{subfigure}
    \caption{(a) Numerical (symbols) and simulation (lines) results for the variation of $\langle W_\mathrm{out}\rangle$ with $r$ and in the TPSS, for $k_0=2$ (black), 5 (red), 10 (blue) and 15 (magenta); the other parameters have the same values as in Fig.~\ref{fig:TPSS_sigma}. The numerical results are obtained using Eq.~\eqref{eq:Wout_explicit}. (b) Variation of engine efficiency $\eta$, given by Eq.~\eqref{eq:eta}, with $r$ and in the TPSS. The heat absorbed $\langle Q_H(\mathcal{T}/2)\rangle$ has been obtained by using Eq.~\eqref{eq:QH}. The inset shows the saturation of the engine efficiency $\eta$ at large $r$ for $k_0=2$, for $T_C=0.1$ (purple) and 0.8 (green); the other parameters have the same values as in Fig.~\ref{fig:TPSS_sigma}.}
    \label{fig:W_out_r}
\end{figure}

Next, we use the correct definition of the extracted work, namely, $W_\mathrm{out}$, in order to compute the efficiency of the engine. In Fig.~\ref{fig:W_out_r}(a), the variation of $\langle W_\mathrm{out}(\mathcal{T})\rangle$ with the resetting rate $r$ has been shown for different values of the stiffness parameter $k_0$. The average work output $\langle W_\mathrm{out}(\mathcal{T})\rangle$ is observed to show a clear non-monotonic dependence on $r$ for $k_0=2$ and $k_0=5$ for the chosen set of parameter values. The non-monotonicity becomes less prominent for $k_0=10$, while it completely disappears for $k_0=15$.  Figure~\ref{fig:W_out_r}(b) shows the corresponding variations in the efficiency, computed using Eq.~\eqref{eq:eta}, with heat absorbed given by Eq.~\eqref{eq:QH}.  The efficiency is found to increase with $r$, but the values are much lower, even for higher values of $r$, as compared to the plot in Fig.~\ref{fig:W_tot_r}. The value of $\eta$ tends to saturate to a value $\lesssim 0.5$ asymptotically with $r$ (see inset of Fig.~\ref{fig:W_tot_r}(b)). On the basis of our results, we infer that $\langle W_\mathrm{out}(\mathcal{T})\rangle$ is the correct definition for the extracted work.

\begin{figure}
    \centering
    \begin{subfigure}{0.9\linewidth}
        \includegraphics[width=\linewidth]{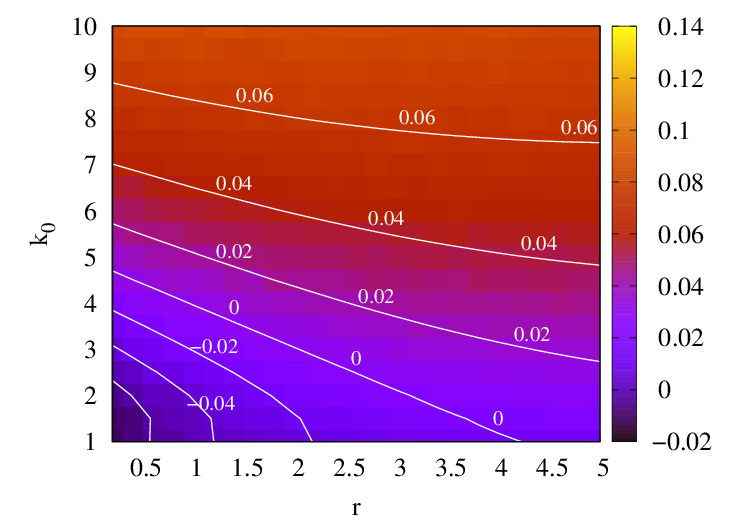}
    \end{subfigure}
    \begin{subfigure}{0.9\linewidth}
        \includegraphics[width=\linewidth]{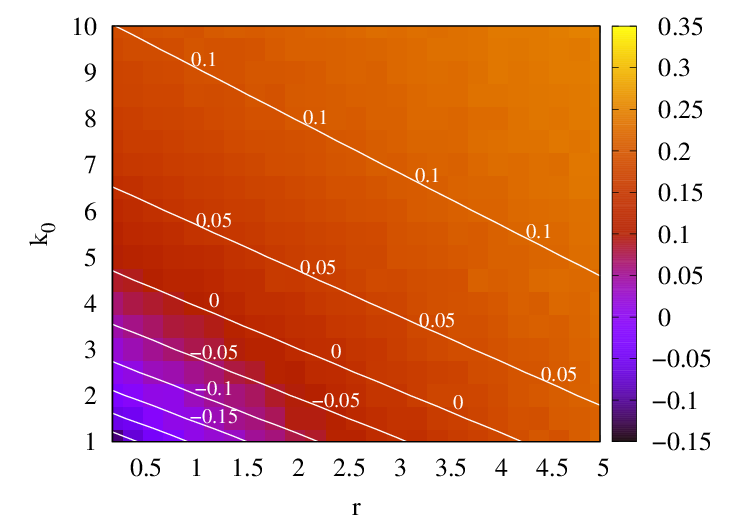}
    \end{subfigure}
        
    \caption{Density plots showing variation of (a) $\langle W_\mathrm{out}\rangle$ and (b) $\eta$ with $r$ and $k_0$ in the TPSS. The white lines show contours with constant values along them as indicated. Other parameters are $T_H=1$ and $T_C=0.5$. The data are generated using the analytical results discussed in the text.}
    \label{fig:PhasePlots}
\end{figure}

\paragraph*{Density plot for efficiency:} In Fig.~\ref{fig:PhasePlots}(a), we show the density plot for the variation of the output work $\langle W_\mathrm{out}(\mathcal{T})\rangle$ with $r$ and $k_0$, while in Fig.~\ref{fig:PhasePlots}(b), the variation in efficiency $\eta$ with respect to the same parameter values is shown. In both the figures, we have used the analytical results developed in Section~\ref{sec:Variance}.  It goes without saying that for the parameter range where the efficiency is negative, the system does not act in the engine mode. 
In fact, to run in the engine mode, the system must be able to provide work as output and take heat as input from the hot bath. Thus, we need to have both mean work output and mean heat absorbed (during the expansion step) to be positive. If any of these quantities switches sign, the efficiency becomes negative, thereby indicating that the system is no more working as an engine.
Both the density plots have been augmented by means of contour lines, which help in easier resolution of the values in conjunction with the color bar.

\paragraph*{Feasibility of defining effective temperatures:} With our set-up containing a Brownian particle trapped in a harmonic potential with a stiffness that is time-independent and equal to $k_0$, one is tempted to compute the effective temperature from the steady-state distribution. From Eq.~\eqref{eq:analytical_variance}, it is easy to see that the steady-state variance is given by 
\begin{align}
    \sigma_r^{\mathrm{ss}} \equiv \lim_{t\to\infty}\sigma_r(t) = \frac{2k_BT}{2k_0+\gamma r}.
    \label{eq:SteadyStateDistribution}
\end{align}
The steady-state variance in absence of resetting equals $k_BT/k$, as may be seen from Eq.~\eqref{eq:MSD-noreset}. We may then equate the expression in Eq.~\eqref{eq:SteadyStateDistribution} to $k_B T_\mathrm{eff}/k_0$, where $T_\mathrm{eff}$ is the effective temperature, to obtain
\begin{align}
    T_\mathrm{eff} &= \frac{2k_0 T}{2k_0+\gamma r}.
    \label{eq:EffectiveTemperature}
\end{align}
 It can be readily seen from Eq.~\eqref{eq:EffectiveTemperature} that the ratios $T_{C,\mathrm{eff}}/T_{H,\mathrm{eff}}$ and $T_C/T_H$ are the same. Now, if a quasistatic engine cycle is carried out, one can replace $k_0$ in the above expression by $k(t)$. This makes the effective temperatures time-dependent, but their ratio still remains equal to $T_C/T_H$. Comparing this value with Fig. \ref{fig:W_out_r}, we infer that the values of $(1-T_{C,\mathrm{eff}}/T_{H,\mathrm{eff}})$ are inconsistent with the obtained efficiency values of our engine. 

\begin{figure}
    \centering
    \includegraphics[width=0.45\textwidth]{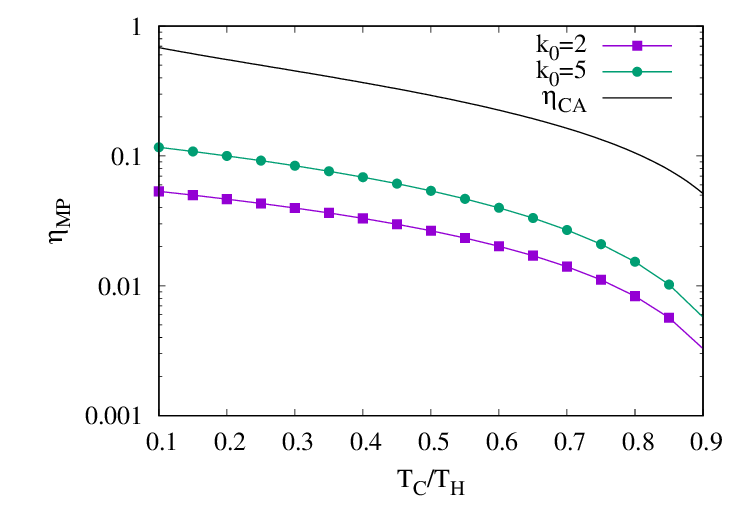}
    \caption{Plot showing the efficiency at maximum power, $\eta_\mathrm{MP}$, as a function of $T_C/T_H$ on a semi-log scale, for $k_0=2$ and $k_0=5$. The other parameters have the same values as in Fig.~\ref{fig:TPSS_sigma}. The values of $\eta_\mathrm{CA}$, the so-called Curzon-Ahlborn efficiency, have been provided (black solid line) for reference.}
    \label{fig:EfficiencyAtMaximumPower}
\end{figure}

\paragraph*{Efficiency at maximum power:} We next proceed to find the efficiency at maximum power, $\eta_\mathrm{MP}$, which is defined as the efficiency of the engine when the output work or power is maximum. We have performed this maximization with respect to the resetting rate $r$. As an example, in Fig.~\ref{fig:W_out_r}(a), the value of $\eta$ computed at the value of $r$ that yields the peak in the curve will be the value of $\eta_\mathrm{MP}$. A heat engine working in the linear response regime yields the value of $\eta_\mathrm{MP}$ given by the \textit{Curzon-Ahlborn efficiency} \cite{cur75_ajp,Schmeidl2008}
\begin{equation}
    \eta_\mathrm{CA} = 1-\sqrt{\frac{T_C}{T_H}}.
    \label{eq:CurzonAhlborn}
\end{equation}
The underlying assumption lies in considering the heat flux to be proportional to the difference in temperature between the heat bath and the system \cite{cur75_ajp}. In the current scenario, the system does not have a well-defined temperature in presence of the nonequilibrium drive due to resetting. However, the analytical techniques developed in~\cite{Schmeidl2008} can be used in this case, where it has been shown that the ensuing leading-order corrections to Eq.~\eqref{eq:CurzonAhlborn} are $\sim \mathcal{O}([\Delta T/T_C]^3)$, with $\Delta T\equiv T_H-T_C \ll T_C$, i.e., $T_C/T_H \gg 1/2$.
In Fig.~\ref{fig:EfficiencyAtMaximumPower}, we have plotted the values of $\eta_\mathrm{MP}$ for our system as a function of $T_C/T_H$. We have considered only the cases where the non-monotonicity has been observed within the range $r\in [1,20]$. Nevertheless, the curve clearly shows that even in the range where $T_C/T_H>1/2$, there is a significant deviation from $\eta_\mathrm{CA}$, the latter being shown as the black solid line in the figure. It can thus be inferred that resetting events drive the engine out of the linear response regime.

\subsection{Reset point not being at the minimum of the potential}

Till now, we have set the reset point to be at the minimum of the potential, namely $x_\mathrm{r}=0$. However, in general, the reset point can be chosen to have a different value that does not correspond to the minimum of the potential. In this case, we generate our results by means of simulations.
\begin{figure}[h!]
    \centering
    \includegraphics[width=0.45\textwidth]{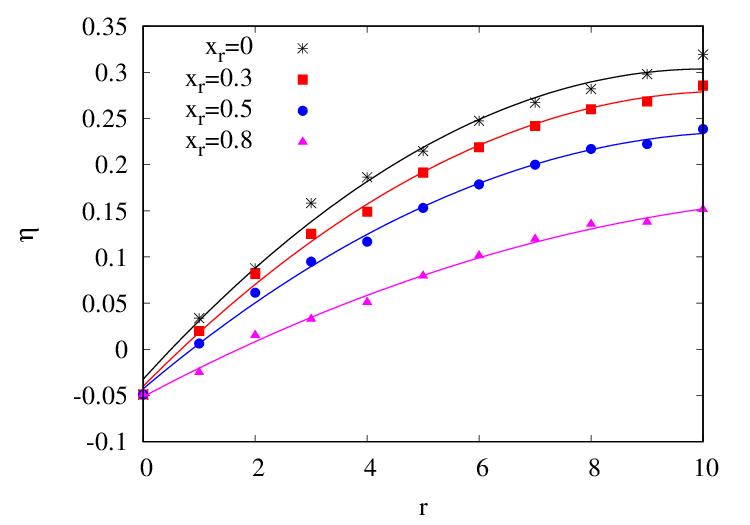}
    \caption{Plots showing the variation of $\eta$ with $r$, for different values of the reset point $x_\mathrm{r}$, for $k_0=2$. The results of simulations are depicted by symbols, while the solid lines are the corresponding quadratic fits. Values of the parameters are the same as in Fig.~\ref{fig:TPSS_sigma}.}
    \label{fig:eta_r_alpha}
\end{figure}
In Fig.~\ref{fig:eta_r_alpha}, we have shown the different plots of efficiency as a function of the resetting rate for different values of $x_\mathrm{r}$. The symbols are the data obtained from simulations, while the solid lines are the quadratic fits. As may be observed,  the higher the value of $x_\mathrm{r}$, the less efficient is the engine at a fixed $r$. The reason is easy to comprehend. 
When the reset point is fixed at the minimum of the potential ($x_\mathrm{r}=0$), the value of $\partial V/\partial t$ is smaller in general. This is because the particle stays close to the minimum of the potential, close to which the variations in the value of $V(x,t)$ with time are smaller as compared to the steeper parts. This increases the average extracted work $\langle W_\mathrm{out}\rangle$ (see Eq.~\eqref{eq:Wout}). However, for $x_\mathrm{r}\ne 0$, the value of $\langle W_\mathrm{out}\rangle$ will be smaller, since $\partial V/\partial t$ is now larger around the reset point. 
Thus, we indeed expect a decrease in efficiency with an increase in $x_\mathrm{r}$. 

\section{Conclusions}
\label{sec:conclusions}

In this work, we addressed an hitherto unexplored theme in the respective fields of stochastic resetting and stochastic heat engines, namely, how does the efficiency of the engine improve on incorporating resetting in the dynamical evolution of the working substance. To this end,  
we incorporated the effect of stochastic resetting into the dynamics of a stochastic engine. The working system consists of a Brownian particle in a harmonic trapping potential with a time-dependent stiffness coefficient, which undergoes repeated resetting at exponentially-distributed random time intervals to a predefined location. An analysis of the working of the engine allowed us to clarify subtle issues related to the identification of suitable thermodynamic quantities quantifying the efficiency of the engine, thereby providing valuable insights into the thermodynamics of systems undergoing resetting. 
Using a renewal equation formalism, we provided formal analytical expressions for the variance of the location of the Brownian particle, which led to an exact formal expression for the work. 
A non-monotonic variation of output work with the rate of resetting was observed for smaller values of trap stiffness, allowing the provision for one to adjust the parameters so as to make the engine run at maximum power. The efficiency, however, exhibited a monotonic growth with $r$. Density plots showing the dependence of the output work and the efficiency on the resetting rate and trap stiffness have been provided. We unveiled that a description of the system in terms of an effective temperature is untenable. The effects of resetting are shown to drive the system away from the linear response regime, by comparing the efficiency at maximum power with the well-known form of the Curzon-Ahlborn efficiency.   
Finally, the consequences of shifting the reset point away from the trap minimum have been explored. In accordance with our expectations, the efficiency decreases with an increase in the distance between the reset point and the trap minimum. Our main conclusion is that the introduction of resetting does play the desirable role in enhancing the performance of the engine when the reset point is close to the minimum of the confining potential of the working substance. As mentioned earlier, such a system can be realized using the techniques of~Refs.~\cite{Ganapathy2021,pal2020experimental}, where a second flashing optical trap can be used for the reset operation, provided the intensity of the associated laser is high enough. Consequently, our observations are amenable to experimental verifications. 

\section*{Acknowledgements} We dedicate this work to the loving memory of Arun M. Jayannavar. SG acknowledges support from the Science and Engineering Research Board (SERB), India under SERB-CRG scheme Grant No. CRG/2020/000596. He also thanks ICTP–Abdus Salam International Centre for Theoretical Physics, Trieste, Italy, for support under its Regular Associateship scheme. SG would like to thank the Isaac Newton Institute for Mathematical Sciences, Cambridge, for support and hospitality during the programme ``Mathematics of movement: An interdisciplinary approach to mutual challenges in animal ecology and cell biology," where work on this paper was completed. This work was supported by EPSRC grant no EP/R014604/1.


\end{document}